\documentclass[10pt,aps,prl,twocolumn]{revtex4-1}
\usepackage{graphicx}
\usepackage{amssymb}
\usepackage{amsmath}

\begin{document}
\title{Easy-Plane Magnetic Strip as a Long Josephson Junction}

\author{Daniel Hill, Se Kwon Kim, and Yaroslav Tserkovnyak}
\affiliation{Department of Physics and Astronomy, University of California, Los Angeles, California 90095, USA}

\begin{abstract}
Spin-torque-biased magnetic dynamics in an easy-plane ferromagnet (EPF) is theoretically studied in the presence of a weak in-plane anisotropy. While this anisotropy spoils U(1) symmetry thereby quenching the conventional spin superfluidity, we show that the system instead realizes a close analog of a long Josephson junction (LJJ) model. The traditional magnetic-field and electric-current controls of the latter map respectively onto the symmetric and antisymmetric combinations of the out-of-plane spin torques applied at the ends of the magnetic strip. This suggests an alternative route towards realizations of superfluid-like transport phenomena in insulating magnetic systems. We study spin-torque-biased phase diagram, providing an analytical solution for static multidomain phases in the EPF. We adapt an existing self-consistency method for the LJJ to develop an approximate solution for the EPF dynamics. The LJJ-EPF mapping allows us to envision superconducting circuit functionality at elevated temperatures. The results apply equally to antiferromagnets with suitable effective free energy in terms of the N\'{e}el order instead of magnetization.
\end{abstract}

\maketitle

\emph{Introduction.}|It has been suggested \cite{soninJETP78,*SoninAP2010,konigPRL01} that insulating thin-film easy-plane ferromagnets (EPF) can exhibit features of superfluid spin transport, which is attractive for spintronics applications, due to low dissipation and long-ranged signal propagation \cite{TakeiPRL2014, TakeiPRL2015}. However, complications arise in that the spin supercurrents, i.e., spin transport with topologically-suppressed dissipation \cite{soninJETP78}, can be inhibited in an EPF by the presence of magnetic anisotropy within the easy-plane. This spoils the requisite $U(1)$ symmetry and pins the magnetization along a particular direction. Such symmetry-breaking anisotropies always exist in real materials, due to, e.g., underlying crystal symmetries or shape anisotropy, demoting the spin superfluid analogy to an imperfect one.

In this Letter, departing from the previous view of the EPF with in-plane anisotropy as a defective spin superfluid, we propose to describe it as a magnetic analog of a long Josephson junction (LJJ), which consists of two superconductors sandwiching a thin insulating layer \cite{OwenPR1967}. This incorporates the in-plane anisotropy as a natural, and, in fact, a potentially desirable ingredient. Specifically, we consider the magnetic dynamics of the EPF driven by the out-of-plane spin torques exerted at its ends. The mapping between EPF and LJJ represents a key result of the paper: Domain walls in the former are translated into phase vortices in the latter, and the symmetric and antisymmetric combinations of the spin torques in the former are translated into the magnetic-field and electric-current controls of the latter. Through the analogy to LJJ, we find a nonequilibrium phase diagram of the EPF, including exact static solutions and certain approximate dynamic solutions. To this end,  we adapt the stability analysis of the static sine-Gordon equation presented in Refs.~\cite{KuplevakhskyPRB2007,KuplevakhskyPRB2006}, along with dynamic solutions of Ref.~\cite{JaworskiSST2008}.

In the following, we construct the spin-torque-biased phase diagram, in which the multivortex stationary states of the LJJ get mapped onto multi-magnetic-domain-wall stationary states in the EPF. The mapping from the equations of motion for the LJJ to the Landau-Lifshitz-Gilbert equations for the EPF is exact for static cases, thus giving the full analytical solution for static multidomain phases in the EPF. For dynamic cases, the equations of motion for the EPF differ from those of the LJJ in that the dissipative leakage at the boundaries due to spin pumping \cite{tserkovPRL02sp} must be accounted for. As this additional feature only changes the boundary conditions, techniques for approximating the dynamical solutions in LJJ's can be carried over to the EPF with minor adjustments. As an example, we develop an approximate analytical solution for the EPF dynamics by adapting an existing self-consistent method for the LJJ \cite{JaworskiSST2008}.

\begin{figure}
	\centering
		\includegraphics[width=1\linewidth]{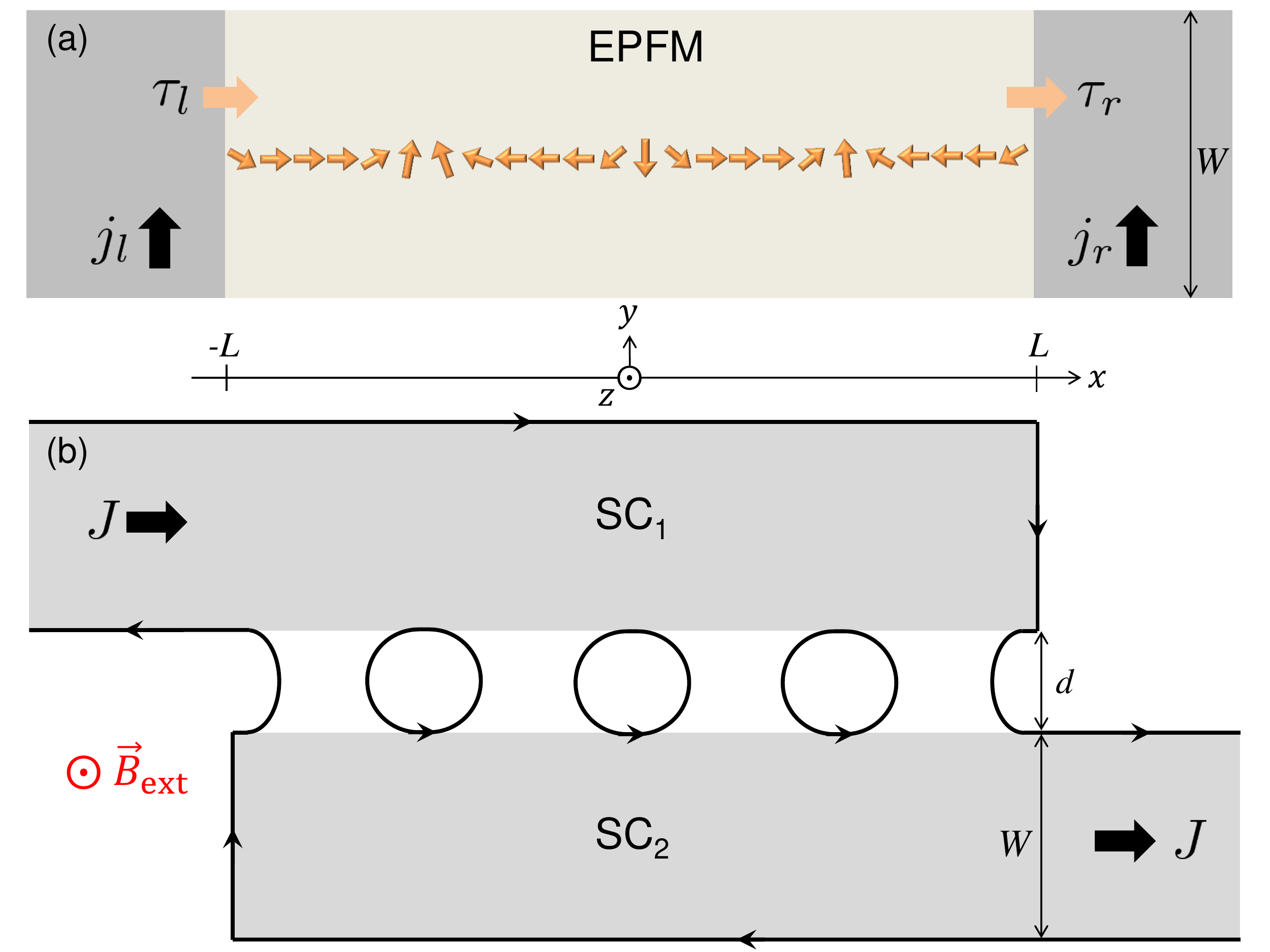}
	\caption{
	Schematics of the magnetic (a) and the original, superconducting (b) long Josephson junctions, depicting domain-wall and vortex configurations, respectively. (a) Easy-plane ferromagnet with spin injection from metal contacts on two side. The spin polarization of the current is along the $z$ axis, while the spin-current flow is oriented along the $x$ axis. The ferromagnet is sufficiently narrow in the transverse dimensions to treat it as quasi-one-dimensional. (b) Diagram of the inline LJJ, with $W$ much smaller than $L$ as well as the depth of the structure in the $z$ direction. We assume two conventional superconductors (SC) much larger than the London penetration depth in all dimensions. The magnetic-field screening currents as well as the circulating vortices are schematically depicted with black oriented lines. The vortices in the junction map onto the domain walls depicted in (a). }
	\label{fig:FMdiagram}
\end{figure}

\emph{Magnetic model.}|In this Letter, we show that LJJ equations of motion can be realized by a magnetic strip connected at its ends to spin-injection leads. We illustrate this by considering a simple structure depicted in Fig.~\ref{fig:FMdiagram}(a). An insulating EPF of length $2L$ is subjected to spin torques $\tau_{r,l}$ applied at its left (right) interface. The underlying spin currents are injected via the spin Hall effect \cite{sinovaRMP15} with spins oriented out of the magnetic easy ($xy$) plane. The system is similar to the easy-plane thin-film ferromagnetic junction of Ref.~\cite{TakeiPRL2014}, but with the addition of a small (compared to the easy-plane, $K$) in-plane anisotropy $K'$. Our magnetic free energy is given by
\begin{equation}
F[\phi,n]=\frac{1}{2}\int d^2 r \left[ A(\partial_\mathbf{r}\phi)^2+Kn^2+K'\sin^2 \phi \right]\,,
\label{fenergy}
\end{equation}
where $\phi(\mathbf{r},t)$ is the azimuthal angle of the directional (unit-vector) order parameter
\begin{equation}
\mathbf{n}(\mathbf{r},t)\equiv(\sqrt{1-n^2}\cos\phi,\sqrt{1-n^2}\sin\phi,n)
\end{equation}
projected onto the $xy$ plane. Its $z$ projection $n(\mathbf{r},t)$ parametrizes the generator of spin rotations in the plane, which thus dictates the Poisson bracket $s\{\phi,n\}=\delta(\mathbf{r}-\mathbf{r}')$ and establishes the canonical conjugacy of the pair $(\phi,sn)$ \cite{halperinPR69}. $s$ is the saturation spin density and $A$ is the order-parameter stiffness. The hard-$z$-axis anisotropy $K\gg K'$ keeps the magnetization dynamics predominantly near the $xy$ plane, which allows us to neglect the gradient terms involving $n$. The ground-state orientation is collinear with the $x$ axis, according to the magnetic anisotropy $\propto K'$, dictating the presence of metastable domain-wall textures, as depicted in Fig.~\ref{fig:FMdiagram}(a).

The dissipation associated with the magnetization dynamics is introduced in the conventional Gilbert-damping form \cite{gilbertIEEEM04}, which for our easy-plane dynamics reduces to the Rayleigh dissipation function (per unit area) of $R=\alpha_d s(\partial_t\phi)^2/2$, parametrized by a damping constant $\alpha_d$. We assume the low-bias regime so as to prevent significant departures of the magnetization away from the easy plane, which corresponds to the limit $|\partial_\mathbf{r} \phi | \ll \sqrt{K/A}$. (This sets the Landau criterion for the stability of planar textures \cite{soninJETP78}.) Thermal nucleation of magnetic vortices responsible for superfluid-like phase slips \cite{kimPRB16} is likewise neglected.

Putting these ingredients together, we obtain the following equations of motion, using the above (effective) Hamiltonian and dissipation functions, $F$ and $R$:
\begin{align}
s\partial_t\phi&=\partial_nF=Kn\,,\\
s\partial_tn&=-\partial_\phi F-\partial_{\partial_t \phi}R=A\partial_\mathbf{r}^2\phi-\frac{K'}{2}\sin2\phi-\alpha_d s\partial_t\phi\,.
\label{dtn}
\end{align}
The remaining issues concern the boundary conditions for spin injection/pumping. The total out-of-plane spin-current densities through the right (left) interface, in the positive $x$ direction, are given by \cite{TakeiPRL2014}
\begin{equation}
j^{(s)}_{r,l} = \mp\frac{g}{4\pi}\left[\mu^{(s)}_{r,l} - \hbar\partial_t\phi\right]\,,
\end{equation}
where $g$ is the (real part of) the spin-mixing conductance of the interface (per unit length) and $\mu^{(s)}$ is the out-of-plane spin accumulation near the interface, which is induced by the spin Hall effect in the metal contacts. For the sake of simplicity, we assume the spin-mixing conductances to be the same for both interfaces. Recognizing the stiffness $\propto A$ term on the right-side of Eq.~\eqref{dtn} as stemming from the bulk spin current $\mathbf{j}^{(s)}=-A\partial_\mathbf{r}\phi$, so that $s\partial_t n=-\partial_\mathbf{r}\cdot\mathbf{j}^{(s)}+\dots$ \cite{soninJETP78}, we invoke spin continuity to obtain the boundary conditions:
\begin{equation}
-A\partial_x\phi(\pm L,t)=j^{(s)}_{r,l}=\tau_{r,l} \pm \gamma\partial_t\phi(\pm L,t) \, .
\label{FMbc}
\end{equation}
Here,
\begin{equation}
\tau_{r,l}\equiv\mp\frac{g}{4\pi}\mu^{(s)}_{r,l}=\frac{\hbar\tan\theta_{\rm SH}}{2e}j_{r,l}
\end{equation}
is the spin Hall torque at the left (right) interface generated by an electric current density $j_{r,l}$ flowing in the $y$ direction through the metal leads. $\theta_{\rm SH}$ is the effective spin Hall angle of the interfaces \cite{tserkovPRB14}. $\gamma \equiv \hbar g / 4\pi$ parametrizes spin pumping out of the ferromagnet by the magnetic dynamics \cite{tserkovPRL02sp}.

Eliminating $n$ from the equations of motion and applying the substitution
\begin{equation}
\tilde{\phi}=2\phi \, ,
\label{phimap}
\end{equation}
we arrive at the damped sine-Gordon equation:
\begin{equation}
\partial_x^2 \tilde{\phi}=\frac{\partial_t^2 \tilde{\phi}}{u^2}+\frac{\sin\tilde{\phi}}{\lambda^2}+\beta_d \partial_t \tilde{\phi} \, ,
\label{SG}
\end{equation}
with the wave speed $u=\sqrt{AK}/s$, characteristic domain-wall width $\lambda=\sqrt{A/K'}$, and damping constant $\beta_d=\alpha_d s/A$. This equation admits a solution of an isolated domain wall as well as low-amplitude spin-wave solutions which obey the massive Klein-Gordon equation, with the mass proportional to $K'$. In the large spin-current limit, so that $|\partial_\mathbf{r} \phi | \gg 1/\lambda$, the excitations become approximately massless. In this (linearly-dispersing) limit, the system approaches the behavior of the EPF without in-plane anisotropy, thus allowing for states that closely resemble the spin superfluid of Ref.~\cite{TakeiPRL2014}. 

The damped sine-Gordon equation has found application in a number of disciplines \cite{sineGordon2014}. The equation is commonly studied in relation to its physical realizations in coupled series of pendulums and long Josephson Junctions. Below we exploit some relevant results of the latter.

\emph{Relation to long Josephson junctions.}|It is instructive to recall the dynamics of the inline configuration of a LJJ \cite{gross2016applied,Parmentier1993}, a diagram of which is depicted in Figure~\ref{fig:FMdiagram}(b). A Josephson junction permits coherent supercurrent tunneling through the insulating region up to a critical current density $j_c$, which depends on the tunneling strength and the superfluid density in the superconductor. In the presence of a magnetic field $\mathbf{B}=B(x)\mathbf{z}$ inside of the junction, we can choose a gauge $\mathbf{A}=A(x,y)\mathbf{x}$, so that $B=-\partial_yA$. The DC Josephson relation for the tunneling current flowing from SC$_1$ to SC$_2$ is then
\begin{equation}
j=j_c\sin\vartheta+gV\,,
\label{jc}
\end{equation}
where $\vartheta(x)=\theta_1-\theta_2$ is the superconducting phase difference across the junction, and we also added the normal current component proportional to the conductance (per unit area) $g$ and the local voltage $V(x)=V_1-V_2$ across the junction. The current enters into the Amp\`ere-Maxwell equation
\begin{equation}
\partial_x B=\frac{4\pi}{c}j+\frac{\varepsilon}{c}\partial_tE\,,
\end{equation}
where $E=-\mathbf{E}\cdot\mathbf{y}=V/d$ is the electric field in and $\varepsilon$ the permittivity of the insulating region.

Next, we invoke the superconducting phase evolution equation ($e>0$)
\begin{equation}
V=\frac{\hbar}{2e}\partial_t\vartheta
\end{equation}
and the relation $\partial_x\theta=-(2e/\hbar c)A$ well inside of the superconducting regions (on the scale of the London penetration depth $\lambda_L$), where the supercurrent vanishes, which leads to
\begin{equation}
B=\frac{A_2-A_1}{d+2\lambda_L}=\frac{\hbar c}{2e(d+2\lambda_L)}\partial_x\vartheta\,.
\label{B}
\end{equation}

Putting Eqs.~\eqref{jc}-\eqref{B} together, we finally reproduce the damped sine-Gordon equation \eqref{SG}, thus identifying the Swihart velocity $u = c\sqrt{\frac{d}{\varepsilon (d+2\lambda_L)}}$, the Josephson penetration depth $\lambda_J = c\sqrt{\frac{\hbar }{8\pi e (d+2\lambda_L) j_c}}$, and the damping parameter $\beta_d=\frac{4\pi g(d+2\lambda_L)}{c^2}$.

The boundary conditions are obtained from Eq.~\eqref{B} by noting that
\begin{equation}
B(\pm L)=B_{\rm ext}\pm\frac{2\pi}{c}J\,,
\end{equation}
where $B_{\rm ext}$ is the externally applied field in the $z$ direction and $J$ is the applied current through the system, per unit of length in the $z$ direction. Comparing this with Eq.~\eqref{FMbc}, we see that the symmetric (antisymmetric) combination of the torques, $\tau_r\pm\tau_l$, realizes the effect of the external field $B_{\rm ext}$ (applied current $J$), in the mapping from the LJJ to the EPF:
\begin{equation}
\tau_{r,l}\rightleftharpoons-\frac{e}{\hbar c}A(d+2 \lambda_L) \left( B_{\text{ext}} \pm \frac{2\pi}{c}J \right) \, .
\end{equation}
The equations of motion of the LJJ and anisotropic EPF systems differ only in the addition of the boundary spin pumping term, $\gamma \partial_t\phi$, in Eq.~(\ref{FMbc}). If spin pumping is negligible (compared to the bulk damping), the two problems are equivalent, with the domain-wall width $\lambda$ replacing the Josephson penetration depth $\lambda_J$ as the key length scale for the order-parameter textures. In particular, the symmetric torque $\tau_r=\tau_l$ injects static domain-wall textures into the EPF, while the external field $B_{\rm ext}$ produces a static multivortex configuration, in the absence of the current $J$.

\begin{figure}
	\centering
		\includegraphics[width=0.45\textwidth]{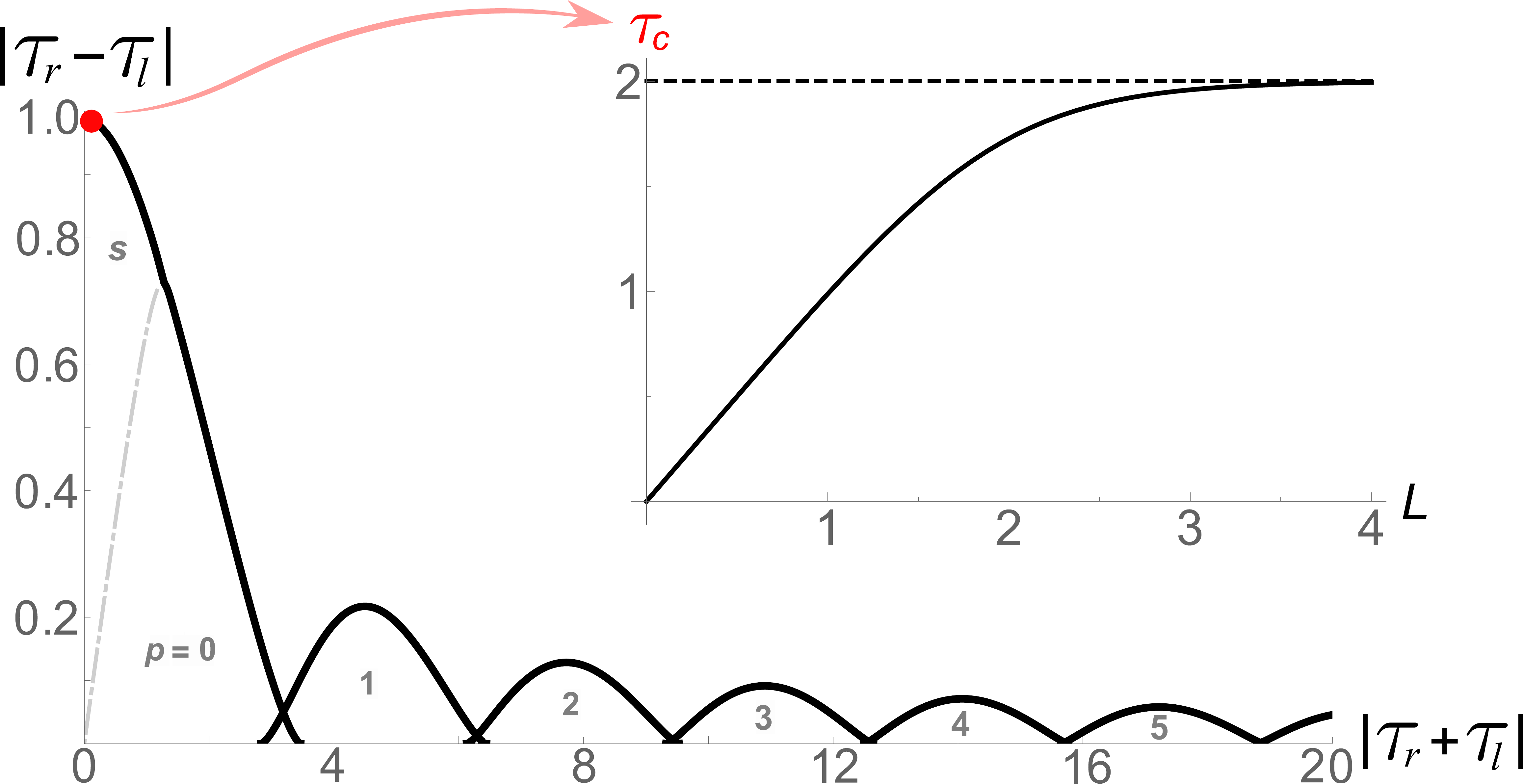}
	\caption{The regions of stability for equalibium $p$-vortex solutions of the LJJ boundary-value problem, for $L/\lambda=1$. The spin torques $\tau_{r,l}$ are in units of $A/\lambda$. Overlapping regions can have either solution. Outside of these regions, i.e., in the high $|\tau_r-\tau_l|$ limit, there are no stable time-independent solutions. Inset: The dependence of the critical torque $\tau_c$, in the asymmetric case, $\tau_r+\tau_l=0$, on the dimensionless length $L$.}
	\label{fig:JvsHstability}
\end{figure}

Having mapped the equations between the models of the LJJ and EPF, for time-independent solutions, allows the equilibrium stability analysis of Ref.~\cite{KuplevakhskyPRB2007} to carry over. The close analogy of the EPF system to the thoroughly studied LJJ model allows us to immediately draw several conclusions about the static solutions. The substitution (\ref{phimap}) indicates that a $2\pi$ phase vortex in the LJJ model corresponds to a domain wall ($\pi$ rotation) in the EPF. The number of stable vortices (or domain walls) is dependent on the boundary conditions and the solution for a given boundary condition can be multivalued, resulting in hysteretic effects. The multivalued solutions are dependent on the length of the system, in units of $\lambda$. Figure~\ref{fig:JvsHstability} shows the regions of stability for $p$-vortex equilibrium solutions for the case of $L=\lambda$. The stability regions have greater overlap in the limit of large $L/\lambda$, and in this limit the edge of the $p>0$ stability region asymptotically approaches the zero bias point as $\sim e^{-L/p\lambda}$. See Ref.~\cite{KuplevakhskyPRB2007} for the analytic equations for computing the phase boundaries.

\emph{Analytic equilibrium solutions.}|Using the stable solutions of the LJJ problem studied in Ref.~\cite{KuplevakhskyPRB2007}, we can map back to the EPF to find the static domain-wall configurations. The full details of the solutions are too lengthy to include here. We give the form of the $\phi(x)$ solutions after mapping to the EPF, as well as some general remarks, and refer to Refs.~\cite{KuplevakhskyPRB2006, KuplevakhskyPRB2007} for further details. From the $p$-vortex LJJ solutions, we find $p$-domain-wall solutions in the EPF have the form 
\begin{equation}
\begin{split}
\phi(x)=\eta
\begin{cases}
\frac{\pi}{2}(p-1)+\text{am}\left(\xi+K(k),k\right)\,, &\text{for } p\text{ even}\\ 
\frac{\pi}{2}p+\text{am}\left(\xi,k\right)\,, &\text{for } p\text{ odd}
\end{cases}\,,
\end{split}
\label{eq:ansol}
\end{equation}
where $\xi=\frac{x}{k\lambda}+\alpha$, and $\eta=\pm1$ for $\tau_r+\tau_l\lessgtr0$. Here, $\alpha$ and $k$ are parameters determined by the boundary conditions \cite{KuplevakhskyPRB2007}, and $\text{am}(u)$ and $K(u)$ are the Jacobi amplitude function and complete elliptic integral of the first kind, respectively.

The zero-domain-wall region includes a portion, separated by the gray line and labeled by $s$ in Fig.~\ref{fig:JvsHstability}, in which Eq.~\eqref{eq:ansol} no longer holds. The solution in the $s$ region has the form 
\begin{equation}
\phi(x)= \zeta \cos^{-1}\left[ k\frac{\text{cn} \left(x/\lambda+\beta\right) }{\text{dn} \left(x/\lambda+\beta\right) } \right]\, ,
\label{ssol}
\end{equation}
where $\beta$ is another parameter determined by the boundary conditions, $\zeta=\pm1$ for $\tau_r-\tau_l\lessgtr0$, and $\text{cn}(u)$ and $\text{dm}(u)$ are the Jacobi elliptic cosine and delta amplitude, respectively. In the limit of $L/\lambda \rightarrow \infty$, the $s$ crossover line becomes a phase-transition line from a no-domain-wall phase to a many-domain-wall phase, but away from this limit the crossover from $s$ to $p=0$ is smoothed out by finite-size effects and no phase transition takes place. 

In the special case of perfectly asymmetric boundary conditions, i.e., $\tau_r=-\tau_l$, the equilibrium solution is given by Eq.~\eqref{ssol} with $\beta=0$, up to the critical value of $|\tau_r-\tau_l|\to\tau_c$. This critical asymmetric torque $\tau_c$ is analogous to the critical current $J_c$ in the LJJ model, with, as shown in the Fig.~\ref{fig:JvsHstability} inset, its value depending on the normalized length of the system. $\tau_c$ approaches $A/\lambda$ asymptotically as $L\rightarrow \infty$ and diminishes as $\tau_c=LK'$ for $L\rightarrow 0$. For yttrium iron garnet, $A\sim10^{-11}\;\text{J/m}^2$, so the saturated critical torque (per unit area), corresponding to $\lambda\sim100\;\text{nm}$ would be $A/\lambda\sim 10^{-4} \;\text{J/m}^2$. Using the spin Hall angle $\theta_{\rm SH}\sim0.1$, the corresponding electrical current density that needs to be applied at the metallic contacts in order to approach $\tau_c$ is of order $10^{12}\;\text{A/m}^2$, which is high but feasible.

\emph{Dynamic solutions.}|Here, inspired by the LJJ analogy, we apply a method similar to that of Ref.~\cite{JaworskiSST2008} in finding an approximate dynamic, spin-propagating solution for the EPF equations of motion, Eq.~(\ref{SG}) with boundary conditions (\ref{FMbc}). To simplify the discussion, we adopt dimensionless notation, such that $A=u=\lambda=1$.
It is natural to start with a trial solution of the form
\begin{equation}
\tilde{\phi}(x,t) = \Omega t + f(x) + \epsilon(x,t)\, .
\label{trial}
\end{equation}
where $\epsilon(x,t)$ is a small periodic function with the (yet to be determined) period $T=2\pi/\Omega$ and with zero time average, and $f(x)$ is a to-be-determined time-independent function. We consider the weak in-plane anisotropy limit for which $\epsilon(x,t) \ll 1$. The boundary conditions are 
\begin{equation}
-\partial_x f(\pm L)= 2\tau_{r,l}\pm\gamma\Omega
\end{equation}
and
\begin{equation}
\partial_x \epsilon(\pm L,t)=0\,,
\label{ebc}
\end{equation}
where we discard the boundary term $\gamma\partial_t\epsilon$ by considering the $\gamma \ll 1$ limit \footnote{This approximation is made to simplify the discussion. Without this constraint, $\epsilon$ satisfies $\partial_x\epsilon\pm\gamma\partial_t\epsilon=0$ at the right (left) boundary, and the resulting solution is similar but lengthier than the one presented here.}.
Plugging the trial solution \eqref{trial} into the sine-Gordon equation \eqref{SG} and averaging over the period $T$, denoted by $\left\langle \dots \right\rangle_T$, we get the time-independent equation
\begin{equation}
\partial_x^2 f=\beta_d\Omega + \big\langle\text{sin}\,\tilde{\phi}\big\rangle_T\,.
\end{equation}
Integrating and applying the boundary conditions, we find the self-consistency equation
\begin{equation}
\begin{split}
f(x) = \int_{-L}^x dx_1 \int_{-L}^{x_1} & dx_2  \left[\beta_d\Omega  + \big\langle \text{sin}\,\tilde{\phi}\big\rangle_T(x_2)\right] \\
& -(2\tau_l-\gamma\Omega) x\,,
\end{split}
\label{seco1}
\end{equation}
with the constraint 
\begin{equation}
2L\beta_d\Omega + \int_{-L}^{L} dx \big\langle \text{sin}\,\tilde{\phi}\big\rangle_T=2(\tau_l-\tau_r -\gamma\Omega)\,.
\label{omega}
\end{equation}
Note that the integral on the right-hand side of Eq.~(\ref{seco1}) depends on both $f(x)$ and $\epsilon(x,t)$ through $\tilde{\phi}$. 

\begin{figure}
	\centering
		\includegraphics[width=0.45\textwidth]{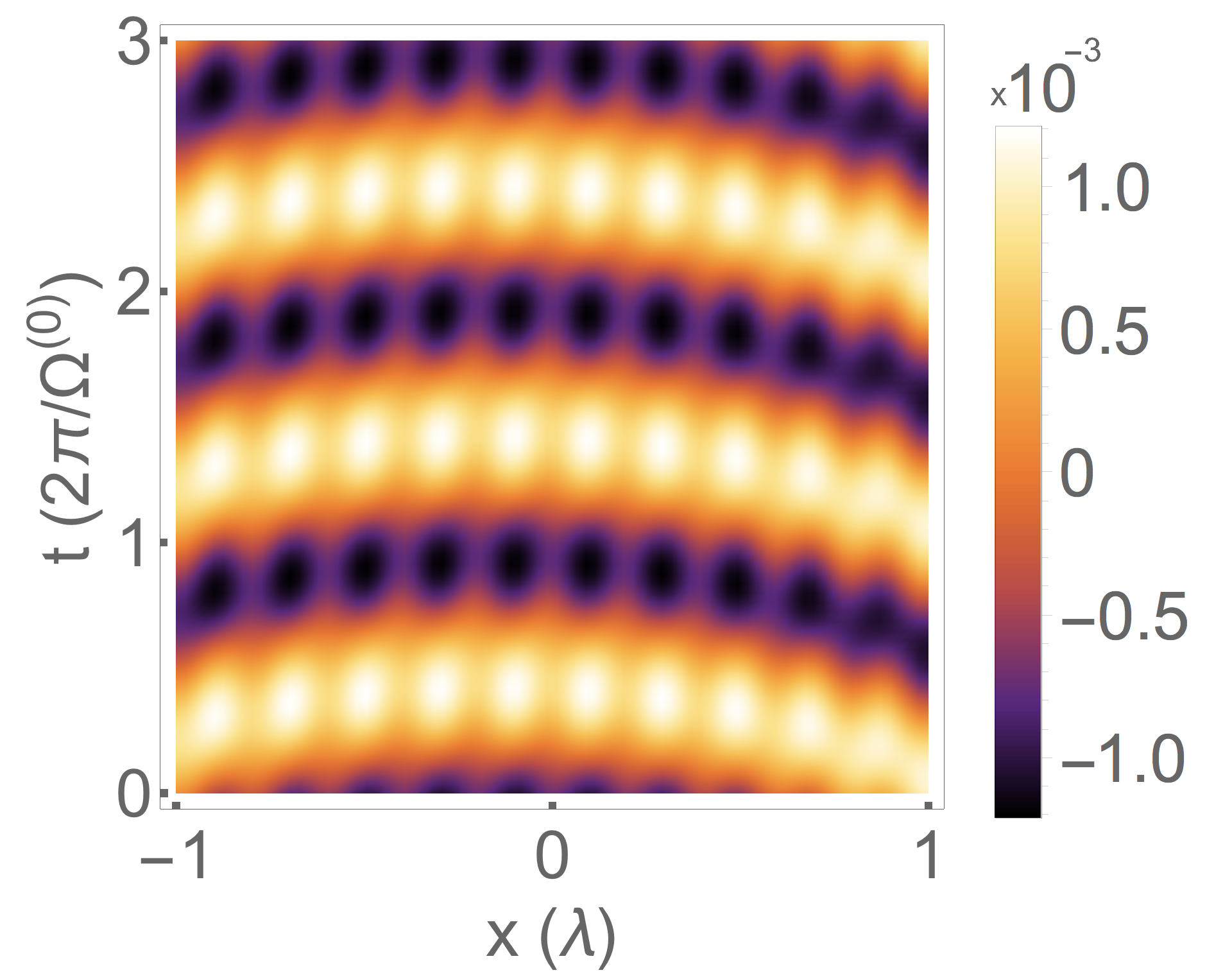}
	\caption{Plot of the approximate modulation of the superfluid phase solution, $\epsilon^{(0)}(x,t)$, resulting from a weak in-plane anisotropy, with $L=1$, $\beta_d=0.1$, $\gamma=0.01$, $\tau_l=1.5$, and $\tau_r=-2$.}
	\label{fig:Epsilon}
\end{figure}

For the time-dependent part of the solution, the dominant contribution is harmonic in $\Omega t$ and obeys
\begin{equation}
\partial_x^2 \epsilon-\partial_t^2 \epsilon-\beta_d \partial_t \epsilon=\text{sin}(\Omega t+ f)\, .
\end{equation}
The solution satisfying the boundary conditions \eqref{ebc} is then readily found to be 
\begin{equation}
\begin{split}
\epsilon(x,t)=\text{Im} \Big( \frac{e^{i\Omega t}}{2i\omega} \big[ 
e^{i\omega x} F^-(x)-e^{-i\omega x} F^+(x)  \\ 
+A\, \text{cos}(\omega x+\omega L) \big] \Big)\,,
\end{split}
\label{seco3}
\end{equation}
where $A= i \left[e^{i\omega L} F^-(L)+e^{-i\omega L} F^+(L)\right]/\text{sin}(2\omega L)$, $\omega^2=\Omega^2-i\beta_d\Omega$, and the functions $F^{\pm}(x)$ are defined as 
\begin{equation}
F^{\pm}(x)=\int_{-L}^x dx_1\, e^{i f(x_1) \pm i\omega x_1}\,.
\end{equation}
Equations (\ref{seco1}), (\ref{omega}), and (\ref{seco3}) form a system of coupled integral equations for $f(x)$, $\Omega$, and $\epsilon(x,t)$. Approximate solutions can be found iteratively by starting with, for example, $\big\langle \text{sin}\, \tilde{\phi}\big\rangle_T^{(0)}=0$, which implies $f^{(0)}(x)=\beta_d\Omega^{(0)} (x+L)^2/2-(2\tau_l-\gamma\Omega^{(0)})(x+L)$, with $\Omega^{(0)} = (\tau_l-\tau_r)/(L\beta_d+\gamma)$. This agrees with the XY-model solution \footnote{Note that our definitions of $\Omega$ and $L$, chosen to more closely adhere to the LJJ references, differ from Ref.~\cite{TakeiPRL2015} by a factor of 2.} of Ref.~\cite{TakeiPRL2015}. This intermediate solution can be plugged into Eq.~(\ref{seco3}) to get $\epsilon^{(0)}(x,t)$, an example of which is plotted in Fig.~\ref{fig:Epsilon}. These in turn can be used to evaluate $\big\langle \text{sin}\, \tilde{\phi}\big\rangle_T^{(1)}$ for generating a new set $f^{(1)}(x)$, $\Omega^{(1)}$, and $\epsilon^{(1)}(x,t)$, and so on. 

Note that the overall frequency of oscillation of the superfluid phase, $\Omega$, is modulated as a function of the EPF length, $L$, as a result of the in-plane anisotropy. This is seen through the dependence of $\Omega$ on the integral on the right-hand side of Eq.~\eqref{omega}. The predicted dependence of $\Omega$ on $L$ can in practice provide a useful experimental probe of the underlying physics.

\emph{Discussion.}|We have revealed an analogy between the spin-torque-biased magnetic dynamics of the EPF with a small in-plane anisotropy and the dynamics of the LJJ. This allowed us to obtain the nonequilibrium phase diagram of multi-domain-wall stationary states and the approximate dynamic solutions by appropriately adapting the existing results for the LJJ. Given that the Josephson junction systems have many potential uses as computing circuit elements, with examples that have been proposed for use in both transistor \cite{PepeSST2001} and memristor \cite{Guarcello2016} constructions, the close analogy between anisotropic easy-plane ferromagnets and LJJ suggests a potential for similar spintronic applications. In principle, the number of closely overlaping stability regions is arbitrary for sufficiently large $L/\lambda$, so one possible application of the anisotropic EPF is a multibit register with memory due to the hysteretic switching between different domain-wall numbers. A significant practical benefit of using magnetic materials instead of superconductors as building blocks of circuit elements is that the relevant physics, such as the spin Hall torque \cite{Miron2011}, is operative at room temperature.

\bibliographystyle{apsrev4-1}
\bibliography{master}

\begin{thebibliography}{23}%
\makeatletter
\providecommand \@ifxundefined [1]{%
 \@ifx{#1\undefined}
}%
\providecommand \@ifnum [1]{%
 \ifnum #1\expandafter \@firstoftwo
 \else \expandafter \@secondoftwo
 \fi
}%
\providecommand \@ifx [1]{%
 \ifx #1\expandafter \@firstoftwo
 \else \expandafter \@secondoftwo
 \fi
}%
\providecommand \natexlab [1]{#1}%
\providecommand \enquote  [1]{``#1''}%
\providecommand \bibnamefont  [1]{#1}%
\providecommand \bibfnamefont [1]{#1}%
\providecommand \citenamefont [1]{#1}%
\providecommand \href@noop [0]{\@secondoftwo}%
\providecommand \href [0]{\begingroup \@sanitize@url \@href}%
\providecommand \@href[1]{\@@startlink{#1}\@@href}%
\providecommand \@@href[1]{\endgroup#1\@@endlink}%
\providecommand \@sanitize@url [0]{\catcode `\\12\catcode `\$12\catcode
  `\&12\catcode `\#12\catcode `\^12\catcode `\_12\catcode `\%12\relax}%
\providecommand \@@startlink[1]{}%
\providecommand \@@endlink[0]{}%
\providecommand \url  [0]{\begingroup\@sanitize@url \@url }%
\providecommand \@url [1]{\endgroup\@href {#1}{\urlprefix }}%
\providecommand \urlprefix  [0]{URL }%
\providecommand \Eprint [0]{\href }%
\providecommand \doibase [0]{http://dx.doi.org/}%
\providecommand \selectlanguage [0]{\@gobble}%
\providecommand \bibinfo  [0]{\@secondoftwo}%
\providecommand \bibfield  [0]{\@secondoftwo}%
\providecommand \translation [1]{[#1]}%
\providecommand \BibitemOpen [0]{}%
\providecommand \bibitemStop [0]{}%
\providecommand \bibitemNoStop [0]{.\EOS\space}%
\providecommand \EOS [0]{\spacefactor3000\relax}%
\providecommand \BibitemShut  [1]{\csname bibitem#1\endcsname}%
\let\auto@bib@innerbib\@empty
\bibitem [{\citenamefont {Sonin}(1978)}]{soninJETP78}%
  \BibitemOpen
  \bibfield  {author} {\bibinfo {author} {\bibfnamefont {E.~B.}\ \bibnamefont
  {Sonin}},\ }\href@noop {} {\bibfield  {journal} {\bibinfo  {journal} {Sov.
  Phys.--JETP}\ }\textbf {\bibinfo {volume} {47}},\ \bibinfo {pages} {1091}
  (\bibinfo {year} {1978})}\BibitemShut {NoStop}%
\bibitem [{\citenamefont {Sonin}(2010)}]{SoninAP2010}%
  \BibitemOpen
  \bibfield  {author} {\bibinfo {author} {\bibfnamefont {E.~B.}\ \bibnamefont
  {Sonin}},\ }\href {\doibase 10.1080/00018731003739943} {\bibfield  {journal}
  {\bibinfo  {journal} {Adv. Phys.}\ }\textbf {\bibinfo {volume} {59}},\
  \bibinfo {pages} {181} (\bibinfo {year} {2010})}\BibitemShut {NoStop}%
\bibitem [{\citenamefont {K{\"o}nig}\ \emph {et~al.}(2001)\citenamefont
  {K{\"o}nig}, \citenamefont {B{\o}nsager},\ and\ \citenamefont
  {MacDonald}}]{konigPRL01}%
  \BibitemOpen
  \bibfield  {author} {\bibinfo {author} {\bibfnamefont {J.}~\bibnamefont
  {K{\"o}nig}}, \bibinfo {author} {\bibfnamefont {M.~C.}\ \bibnamefont
  {B{\o}nsager}}, \ and\ \bibinfo {author} {\bibfnamefont {A.~H.}\ \bibnamefont
  {MacDonald}},\ }\href@noop {} {\bibfield  {journal} {\bibinfo  {journal}
  {Phys. Rev. Lett.}\ }\textbf {\bibinfo {volume} {87}},\ \bibinfo {pages}
  {187202} (\bibinfo {year} {2001})}\BibitemShut {NoStop}%
\bibitem [{\citenamefont {Takei}\ and\ \citenamefont
  {Tserkovnyak}(2014)}]{TakeiPRL2014}%
  \BibitemOpen
  \bibfield  {author} {\bibinfo {author} {\bibfnamefont {S.}~\bibnamefont
  {Takei}}\ and\ \bibinfo {author} {\bibfnamefont {Y.}~\bibnamefont
  {Tserkovnyak}},\ }\href {\doibase 10.1103/PhysRevLett.112.227201} {\bibfield
  {journal} {\bibinfo  {journal} {Phys. Rev. Lett.}\ }\textbf {\bibinfo
  {volume} {112}},\ \bibinfo {pages} {227201} (\bibinfo {year}
  {2014})}\BibitemShut {NoStop}%
\bibitem [{\citenamefont {Takei}\ and\ \citenamefont
  {Tserkovnyak}(2015)}]{TakeiPRL2015}%
  \BibitemOpen
  \bibfield  {author} {\bibinfo {author} {\bibfnamefont {S.}~\bibnamefont
  {Takei}}\ and\ \bibinfo {author} {\bibfnamefont {Y.}~\bibnamefont
  {Tserkovnyak}},\ }\href {\doibase 10.1103/PhysRevLett.115.156604} {\bibfield
  {journal} {\bibinfo  {journal} {Phys. Rev. Lett.}\ }\textbf {\bibinfo
  {volume} {115}},\ \bibinfo {pages} {156604} (\bibinfo {year}
  {2015})}\BibitemShut {NoStop}%
\bibitem [{\citenamefont {Owen}\ and\ \citenamefont
  {Scalapino}(1967)}]{OwenPR1967}%
  \BibitemOpen
  \bibfield  {author} {\bibinfo {author} {\bibfnamefont {C.~S.}\ \bibnamefont
  {Owen}}\ and\ \bibinfo {author} {\bibfnamefont {D.~J.}\ \bibnamefont
  {Scalapino}},\ }\href {\doibase 10.1103/PhysRev.164.538} {\bibfield
  {journal} {\bibinfo  {journal} {Phys. Rev.}\ }\textbf {\bibinfo {volume}
  {164}},\ \bibinfo {pages} {538} (\bibinfo {year} {1967})}\BibitemShut
  {NoStop}%
\bibitem [{\citenamefont {Kuplevakhsky}\ and\ \citenamefont
  {Glukhov}(2007)}]{KuplevakhskyPRB2007}%
  \BibitemOpen
  \bibfield  {author} {\bibinfo {author} {\bibfnamefont {S.~V.}\ \bibnamefont
  {Kuplevakhsky}}\ and\ \bibinfo {author} {\bibfnamefont {A.~M.}\ \bibnamefont
  {Glukhov}},\ }\href {https://doi.org/10.1103/PhysRevB.76.174515} {\bibfield
  {journal} {\bibinfo  {journal} {Phys. Rev. B.}\ }\textbf {\bibinfo {volume}
  {76}},\ \bibinfo {pages} {174515} (\bibinfo {year} {2007})}\BibitemShut
  {NoStop}%
\bibitem [{\citenamefont {Kuplevakhsky}\ and\ \citenamefont
  {Glukhov}(2006)}]{KuplevakhskyPRB2006}%
  \BibitemOpen
  \bibfield  {author} {\bibinfo {author} {\bibfnamefont {S.~V.}\ \bibnamefont
  {Kuplevakhsky}}\ and\ \bibinfo {author} {\bibfnamefont {A.~M.}\ \bibnamefont
  {Glukhov}},\ }\href {https://doi.org/10.1103/PhysRevB.73.024513} {\bibfield
  {journal} {\bibinfo  {journal} {Phys. Rev. B.}\ }\textbf {\bibinfo {volume}
  {73}},\ \bibinfo {pages} {024513} (\bibinfo {year} {2006})}\BibitemShut
  {NoStop}%
\bibitem [{\citenamefont {Jaworski}(2008)}]{JaworskiSST2008}%
  \BibitemOpen
  \bibfield  {author} {\bibinfo {author} {\bibfnamefont {M.}~\bibnamefont
  {Jaworski}},\ }\href@noop {} {\bibfield  {journal} {\bibinfo  {journal}
  {Supercond. Sci. Technol.}\ }\textbf {\bibinfo {volume} {21}},\ \bibinfo
  {pages} {065016} (\bibinfo {year} {2008})}\BibitemShut {NoStop}%
\bibitem [{\citenamefont {Tserkovnyak}\ \emph {et~al.}(2002)\citenamefont
  {Tserkovnyak}, \citenamefont {Brataas},\ and\ \citenamefont
  {Bauer}}]{tserkovPRL02sp}%
  \BibitemOpen
  \bibfield  {author} {\bibinfo {author} {\bibfnamefont {Y.}~\bibnamefont
  {Tserkovnyak}}, \bibinfo {author} {\bibfnamefont {A.}~\bibnamefont
  {Brataas}}, \ and\ \bibinfo {author} {\bibfnamefont {G.~E.~W.}\ \bibnamefont
  {Bauer}},\ }\href@noop {} {\bibfield  {journal} {\bibinfo  {journal} {Phys.
  Rev. Lett.}\ }\textbf {\bibinfo {volume} {88}},\ \bibinfo {eid} {117601}
  (\bibinfo {year} {2002})}\BibitemShut {NoStop}%
\bibitem [{\citenamefont {Sinova}\ \emph {et~al.}(2015)\citenamefont {Sinova},
  \citenamefont {Valenzuela}, \citenamefont {Wunderlich}, \citenamefont
  {Back},\ and\ \citenamefont {Jungwirth}}]{sinovaRMP15}%
  \BibitemOpen
  \bibfield  {author} {\bibinfo {author} {\bibfnamefont {J.}~\bibnamefont
  {Sinova}}, \bibinfo {author} {\bibfnamefont {S.~O.}\ \bibnamefont
  {Valenzuela}}, \bibinfo {author} {\bibfnamefont {J.}~\bibnamefont
  {Wunderlich}}, \bibinfo {author} {\bibfnamefont {C.~H.}\ \bibnamefont
  {Back}}, \ and\ \bibinfo {author} {\bibfnamefont {T.}~\bibnamefont
  {Jungwirth}},\ }\href {\doibase 10.1103/RevModPhys.87.1213} {\bibfield
  {journal} {\bibinfo  {journal} {Rev. Mod. Phys.}\ }\textbf {\bibinfo {volume}
  {87}},\ \bibinfo {pages} {1213} (\bibinfo {year} {2015})}\BibitemShut
  {NoStop}%
\bibitem [{\citenamefont {Halperin}\ and\ \citenamefont
  {Hohenberg}(1969)}]{halperinPR69}%
  \BibitemOpen
  \bibfield  {author} {\bibinfo {author} {\bibfnamefont {B.~I.}\ \bibnamefont
  {Halperin}}\ and\ \bibinfo {author} {\bibfnamefont {P.~C.}\ \bibnamefont
  {Hohenberg}},\ }\href@noop {} {\bibfield  {journal} {\bibinfo  {journal}
  {Phys. Rev.}\ }\textbf {\bibinfo {volume} {188}},\ \bibinfo {pages} {898}
  (\bibinfo {year} {1969})}\BibitemShut {NoStop}%
\bibitem [{\citenamefont {Gilbert}(2004)}]{gilbertIEEEM04}%
  \BibitemOpen
  \bibfield  {author} {\bibinfo {author} {\bibfnamefont {T.~L.}\ \bibnamefont
  {Gilbert}},\ }\href@noop {} {\ \textbf {\bibinfo {volume} {40}},\ \bibinfo
  {pages} {3443} (\bibinfo {year} {2004})}\BibitemShut {NoStop}%
\bibitem [{\citenamefont {Kim}\ \emph {et~al.}(2016)\citenamefont {Kim},
  \citenamefont {Takei},\ and\ \citenamefont {Tserkovnyak}}]{kimPRB16}%
  \BibitemOpen
  \bibfield  {author} {\bibinfo {author} {\bibfnamefont {S.~K.}\ \bibnamefont
  {Kim}}, \bibinfo {author} {\bibfnamefont {S.}~\bibnamefont {Takei}}, \ and\
  \bibinfo {author} {\bibfnamefont {Y.}~\bibnamefont {Tserkovnyak}},\ }\href
  {\doibase 10.1103/PhysRevB.93.020402} {\bibfield  {journal} {\bibinfo
  {journal} {Phys. Rev. B}\ }\textbf {\bibinfo {volume} {93}},\ \bibinfo
  {pages} {020402(R)} (\bibinfo {year} {2016})}\BibitemShut {NoStop}%
\bibitem [{\citenamefont {Tserkovnyak}\ and\ \citenamefont
  {Bender}(2014)}]{tserkovPRB14}%
  \BibitemOpen
  \bibfield  {author} {\bibinfo {author} {\bibfnamefont {Y.}~\bibnamefont
  {Tserkovnyak}}\ and\ \bibinfo {author} {\bibfnamefont {S.~A.}\ \bibnamefont
  {Bender}},\ }\href {\doibase 10.1103/PhysRevB.90.014428} {\bibfield
  {journal} {\bibinfo  {journal} {Phys. Rev. B}\ }\textbf {\bibinfo {volume}
  {90}},\ \bibinfo {pages} {014428} (\bibinfo {year} {2014})}\BibitemShut
  {NoStop}%
\bibitem [{\citenamefont {Cuevas-Maraver}\ \emph {et~al.}(2014)\citenamefont
  {Cuevas-Maraver}, \citenamefont {Kevrekidis},\ and\ \citenamefont
  {Williams}}]{sineGordon2014}%
  \BibitemOpen
  \bibinfo {editor} {\bibfnamefont {J.}~\bibnamefont {Cuevas-Maraver}},
  \bibinfo {editor} {\bibfnamefont {P.}~\bibnamefont {Kevrekidis}}, \ and\
  \bibinfo {editor} {\bibfnamefont {F.}~\bibnamefont {Williams}},\ eds.,\ \href
  {https://link.springer.com/book/10.1007/978-3-319-06722-3} {\emph {\bibinfo
  {title} {The sine-Gordon Model and its Applications}}},\ \bibinfo {series}
  {Nonlinear Systems and Complexity}, Vol.~\bibinfo {volume} {10}\ (\bibinfo
  {publisher} {Springer, Cham},\ \bibinfo {year} {2014})\BibitemShut {NoStop}%
\bibitem [{\citenamefont {Gross}\ \emph {et~al.}(2016)\citenamefont {Gross},
  \citenamefont {Marx},\ and\ \citenamefont {Deppe}}]{gross2016applied}%
  \BibitemOpen
  \bibfield  {author} {\bibinfo {author} {\bibfnamefont {R.}~\bibnamefont
  {Gross}}, \bibinfo {author} {\bibfnamefont {A.}~\bibnamefont {Marx}}, \ and\
  \bibinfo {author} {\bibfnamefont {F.}~\bibnamefont {Deppe}},\ }\href
  {https://books.google.com/books?id=4SIzrgEACAAJ} {\emph {\bibinfo {title}
  {Applied Superconductivity: Josephson Effect and Superconducting
  Electronics}}},\ de Gruyter Textbook\ (\bibinfo  {publisher} {Walter De
  Gruyter Incorporated},\ \bibinfo {year} {2016})\BibitemShut {NoStop}%
\bibitem [{\citenamefont {Parmentier}(1993)}]{Parmentier1993}%
  \BibitemOpen
  \bibfield  {author} {\bibinfo {author} {\bibfnamefont {R.}~\bibnamefont
  {Parmentier}},\ }in\ \href {\doibase 10.1007/978-94-011-1918-4_7} {\emph
  {\bibinfo {booktitle} {The New Superconducting Electronics. NATO ASI
  Series}}},\ Vol.\ \bibinfo {volume} {251},\ \bibinfo {editor} {edited by\
  \bibinfo {editor} {\bibfnamefont {H.}~\bibnamefont {Weinstock}}\ and\
  \bibinfo {editor} {\bibfnamefont {R.}~\bibnamefont {Ralston}}}\ (\bibinfo
  {publisher} {Springer},\ \bibinfo {address} {Van Godewijckstraat 30, 3311 GX
  Dordrecht, Netherlands},\ \bibinfo {year} {1993})\ Chap.~\bibinfo {chapter}
  {7}, pp.\ \bibinfo {pages} {221--248}\BibitemShut {NoStop}%
\bibitem [{Note1()}]{Note1}%
  \BibitemOpen
  \bibinfo {note} {This approximation is made to simplify the discussion.
  Without this constraint, $\epsilon $ satisfies $\partial _x\epsilon \pm
  \gamma \partial _t\epsilon =0$ at the right (left) boundary, and the
  resulting solution is similar but lengthier than the one presented
  here.}\BibitemShut {Stop}%
\bibitem [{Note2()}]{Note2}%
  \BibitemOpen
  \bibinfo {note} {Note that our definitions of $\Omega $ and $L$, chosen to
  more closely adhere to the LJJ references, differ from Ref.~\cite
  {TakeiPRL2015} by a factor of 2.}\BibitemShut {Stop}%
\bibitem [{\citenamefont {Pepe}\ \emph {et~al.}(2001)\citenamefont {Pepe},
  \citenamefont {Scaldaferri}, \citenamefont {Parlato}, \citenamefont {Peluso},
  \citenamefont {Granata}, \citenamefont {Russo}, \citenamefont {Rotoli},\ and\
  \citenamefont {Booth}}]{PepeSST2001}%
  \BibitemOpen
  \bibfield  {author} {\bibinfo {author} {\bibfnamefont {G.~P.}\ \bibnamefont
  {Pepe}}, \bibinfo {author} {\bibfnamefont {R.}~\bibnamefont {Scaldaferri}},
  \bibinfo {author} {\bibfnamefont {L.}~\bibnamefont {Parlato}}, \bibinfo
  {author} {\bibfnamefont {G.}~\bibnamefont {Peluso}}, \bibinfo {author}
  {\bibfnamefont {C.}~\bibnamefont {Granata}}, \bibinfo {author} {\bibfnamefont
  {M.}~\bibnamefont {Russo}}, \bibinfo {author} {\bibfnamefont
  {G.}~\bibnamefont {Rotoli}}, \ and\ \bibinfo {author} {\bibfnamefont {N.~E.}\
  \bibnamefont {Booth}},\ }\href
  {http://stacks.iop.org/0953-2048/14/i=12/a=301} {\bibfield  {journal}
  {\bibinfo  {journal} {Supercond. Sci. Technol.}\ }\textbf {\bibinfo {volume}
  {14}},\ \bibinfo {pages} {987} (\bibinfo {year} {2001})}\BibitemShut
  {NoStop}%
\bibitem [{\citenamefont {Guarcello}\ \emph {et~al.}(2016)\citenamefont
  {Guarcello}, \citenamefont {Solinas}, \citenamefont {Di~Ventra},\ and\
  \citenamefont {Giazotto}}]{Guarcello2016}%
  \BibitemOpen
  \bibfield  {author} {\bibinfo {author} {\bibfnamefont {C.}~\bibnamefont
  {Guarcello}}, \bibinfo {author} {\bibfnamefont {P.}~\bibnamefont {Solinas}},
  \bibinfo {author} {\bibfnamefont {M.}~\bibnamefont {Di~Ventra}}, \ and\
  \bibinfo {author} {\bibfnamefont {F.}~\bibnamefont {Giazotto}},\ }\href
  {https://www.nature.com/articles/srep46736} {\bibfield  {journal} {\bibinfo
  {journal} {Sci. Rep.}\ }\textbf {\bibinfo {volume} {7}} (\bibinfo {year}
  {2016})}\BibitemShut {NoStop}%
\bibitem [{\citenamefont {Miron}\ \emph {et~al.}(2011)\citenamefont {Miron},
  \citenamefont {Garello}, \citenamefont {Gaudin}, \citenamefont {Zermatten},
  \citenamefont {Costache}, \citenamefont {Auffret}, \citenamefont {Bandiera},
  \citenamefont {Rodmacq}, \citenamefont {Schuhl},\ and\ \citenamefont
  {Gambardella}}]{Miron2011}%
  \BibitemOpen
  \bibfield  {author} {\bibinfo {author} {\bibfnamefont {I.~M.}\ \bibnamefont
  {Miron}}, \bibinfo {author} {\bibfnamefont {K.}~\bibnamefont {Garello}},
  \bibinfo {author} {\bibfnamefont {G.}~\bibnamefont {Gaudin}}, \bibinfo
  {author} {\bibfnamefont {P.-J.}\ \bibnamefont {Zermatten}}, \bibinfo {author}
  {\bibfnamefont {M.~V.}\ \bibnamefont {Costache}}, \bibinfo {author}
  {\bibfnamefont {S.}~\bibnamefont {Auffret}}, \bibinfo {author} {\bibfnamefont
  {S.}~\bibnamefont {Bandiera}}, \bibinfo {author} {\bibfnamefont
  {B.}~\bibnamefont {Rodmacq}}, \bibinfo {author} {\bibfnamefont
  {A.}~\bibnamefont {Schuhl}}, \ and\ \bibinfo {author} {\bibfnamefont
  {P.}~\bibnamefont {Gambardella}},\ }\href {\doibase 10.1038/nature10309}
  {\bibfield  {journal} {\bibinfo  {journal} {Nature}\ }\textbf {\bibinfo
  {volume} {476}},\ \bibinfo {pages} {189} (\bibinfo {year}
  {2011})}\BibitemShut {NoStop}%
\end{thebibliography}%

\end{document}